# The morpho-topographic and cartographic analysis of the archaeological site Corneşti "Iarcuri", Timiş County, Romania, using computer sciences methods (GIS and Remote Sensing techniques)

**Dorel Micle, Marcel Török-Oance, Liviu Măruia**
West University of Timişoara, România

**ABSTRACT**: The archaeological site Cornesti "Iarcuri" is the largest earth fortification in Romania, made out of four concentric compounds, spreading over 1780 hectares. It is known since 1700, but it had only a few small attempts of systematic research, the fortress gained interest only after the publishing of some satellite images by Google Earth. It is located in an area of high fields and it occupies three interfluves and contains two streams. Our paper contains a geomorphologic, topographic and cartographic analysis of the site in order to determine the limits, the structure, the morphology, the construction technique and the functionality of such a fortification.Our research is based on satellite image analysis, on archaeological topography, on soil, climate and vegetation analysis as a way to offer a complex image, through this interdisciplinary study of landscape archaeology. Through our work we try not to date the site as this objective will be achieved only after completing the systematic excavations which started in 2007, but only to analyze the co-relationship with the environment.

**Introduction**

The archaeological site from Corneşti "Iarcuri[i]" is a fortification made up of four concentric waves of earth that takes up a surface of 1780.5 ha. Due to the fact that the only archaeological evidence are the waves of earth and the surface on which the fortification lies is enormous, the fortification from Corneşti "Iarcuri" has been ignored for a very long time and it is almost unknown in specialized literature [Med93], [MMD06].

Located approximately between the settlements of Corneşti (to the South - West), Orţişoara (to the North - West), Murani (to the South – East) and Seceani (to the North - East), the fortification is in the Eastern vicinity of the European way E671, from where one can see the Western side of the enclosure wave no. 4

The great size of the fortification makes it visible as a whole only from great heights. In this sense, a first aerophotogrammetric analysis was conducted in 1988 by M. Rada, N. Ciochină and D. Manea [RCM89], who for the first time talk about the possible existence of four fortified enclosures.

Between 2005 and 2006 a team of researchers from the West University of Timisoara, the History Department (Dorel Micle, Liviu Măruia) and the Geography Department (Marcel Török-Oance), started a scientific activity of archaeological research on the field. The research was started as a result of identifying, on satellite images with the help of Google Earth (DigitalGlobe images), the entire archaeological complex.

The objective was to establish exactly the lining of the waves from the four enclosures of the fortification, to identify all the possible settlements that have existed through time in this space and to collect archaeological material from the surface that would contribute to a better dating and cultural assigning (MMD06). The team verified on the field the data provided by the satellite images, in the spots in which revealing clues (color, shapes, humidity and vegetation) suggested possible human intervention (Fig. 1)

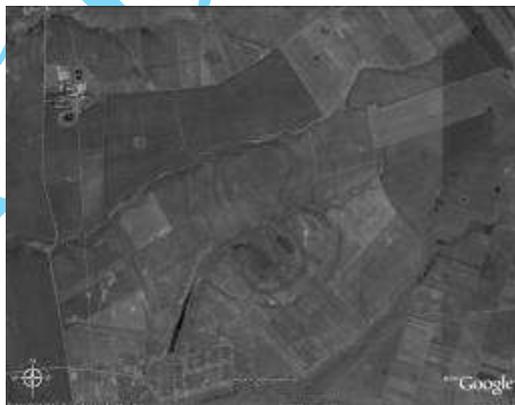

***Fig. 1. Archeological site Corneşti "Iarcuri". Google Earth (DigitalGlobe image)***

The fortification appears on most maps that present the historical and geographical region of Banat, starting with the 18th century, but none of these maps provide a correct image of the size and lining of the archaeological complex of Corneşti.

In the perspective in which the start of systematic archaeological investigations – that would clarify the complex issues that a site of this amplitude raises (chronology, ethnic assignment, function, etc.) – is required, our team of researchers has conducted a preliminary interdisciplinary analysis aiming at corroborating the data provided by the satellite images, air photos, cartography, geomorphology, pedology and computer science.

**1. A Digital Elevation Model**

A Digital Elevation Model – DEM was made both for the morphometric analysis and for building a data base of the spatial values of the studied area. The making of a DEM, starting from the existing maps and topographic plans, supposes the transposition of the altitude values of the topographic map in digital format and the insertion of the existing values by mathematical methods. For the making of this model topographic maps on the scale of 1:25000 were used as data sources. These maps were scanned, referenced geographically in the national coordinate system Stereo 1970 and inlaid. Through manual digitizing of all curve levels and elevation points (with the help of the ArcGIS 9.2 program) the altitude values were extracted and interlaid by the Delauney method.

We used the method of contracting the insertions between the curve levels and we eliminated the errors of insertion generated by the insufficiency of altitude data in some areas with the option "bridge and tunnel edge removal" of the program IDRISI Andes (Eas06). A representation of the relief was thus achieved by the means of a network of irregular triangles known as TIN (Triangular Irregular Network). This was subsequently transformed into a raster model, achieving the numeric model of the field (DEM) with a 2 m resolution (a pixel of the image corresponds to a 2 X 2 m surface on the field). For a better image of the earth waves DEM was improved by using the altitude data gathered on the field (Fig. 2).

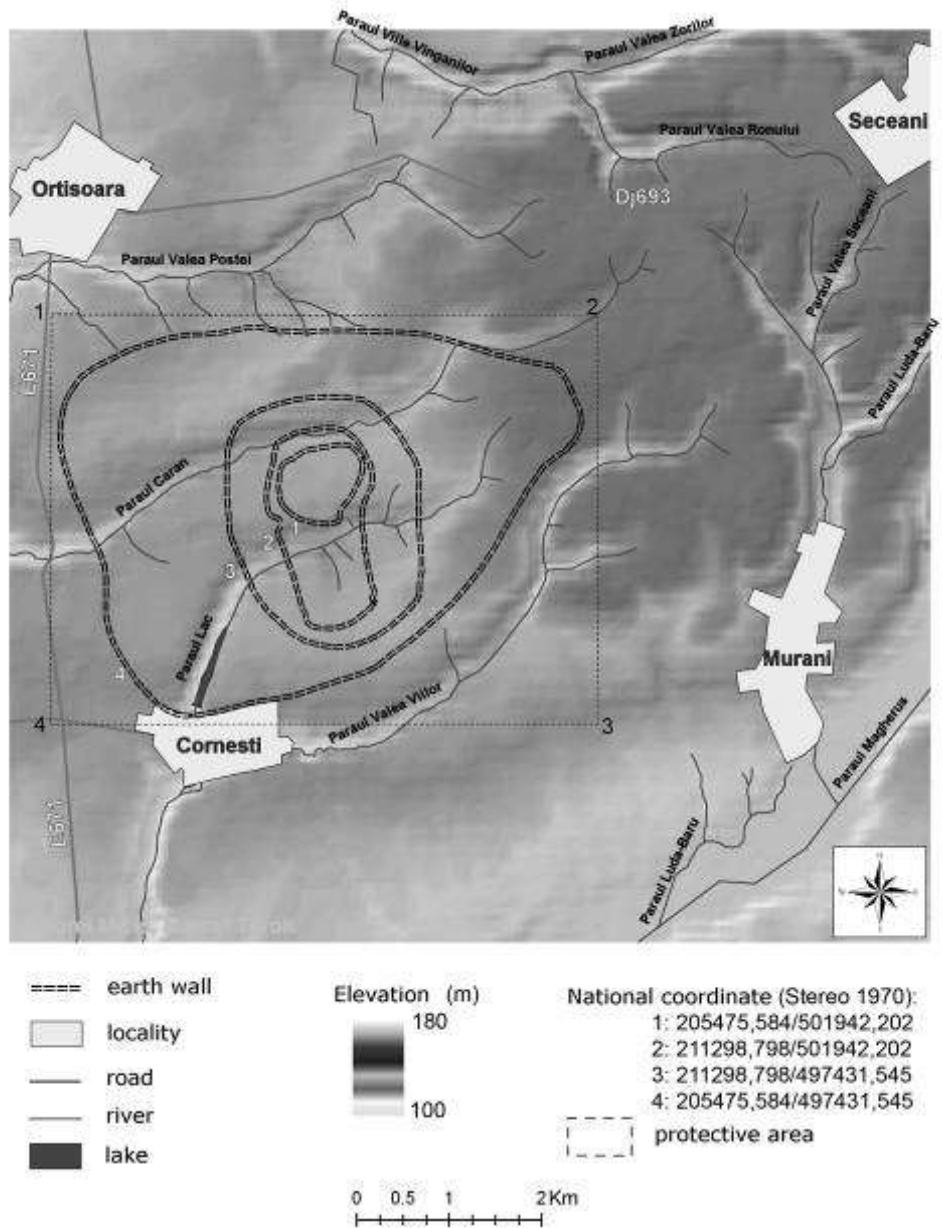

*Fig. 2. Archeological site Corneşti "Iarcuri". Digital Elevation Model*

The other elements of the spatial data base (layers of hydrographic network, the roads, the contour of the earth waves and the living areas) were digitized from the color orthophotoplans because these are the most recent cartographic sources, and the scale of 1:5000, enables the examination of all the details (Fig. 3).

252

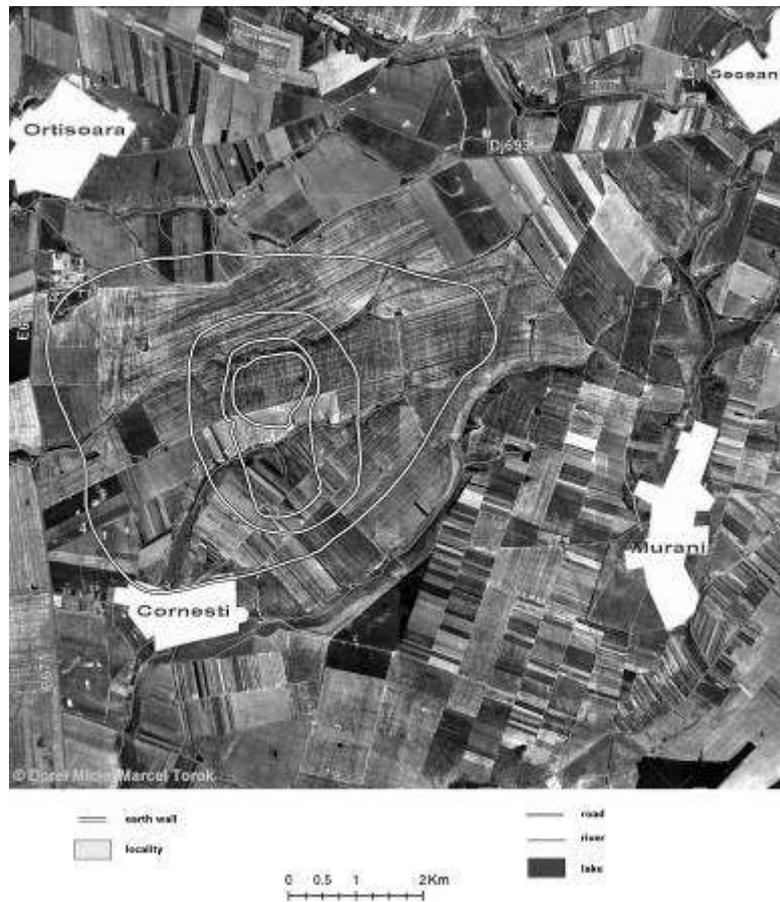

*Fig. 3. Archeological site Corneşti "Iarcuri". Orthophotoplan*

With the help of the DEM a morphometric analysis of the region was made, by analyzing the following morphometric elements: the hypsometry, the slope and the aspect (surface lining). All the morphometric data are synthesized in table 1. DEM was used for the making of 2 topographic profiles (Fig. 4 and 5), one on the lining North - South and the other East – West, and for the visual analysis of the studied region (2D and 3D illustrations).

253

*Table 1. Morphometric data from the DEM for the area in study.*

| Area | The entire area | Enclosure 1 | Enclosure 2 | Enclosure 3 | Enclosure 4 | The site (all enclosures) |
|---|---|---|---|---|---|---|
| **Surface (km$^2$)** | 89,35 | 0,71 | 1,43 | 2,89 | 12,68 | 17,80 |
| **Surface (ha)** | 8935,86 | 71,04 | 143,28 | 289,08 | 1268,18 | 1780,50 |
| **Perimeter (km)** | - | 3,01 | 9,09 | 14,15 | 24,25 | - |
| **Long axis (km)** | - | 1,05 | 2,24 | 2,82 | 5,72 | 5,72 |
| **Short axis (km)** | - | 0,94 | 1,25 | 2,23 | 4,15 | 4,15 |
| **Min. Altitudine (m)** | 97,60 | 134,94 | 125,40 | 158,30 | 117,35 | 117,35 |
| **Max. Altitudine (m)** | 184,65 | 147,51 | 148,04 | 125,24 | 168,56 | 168,56 |
| **Medium Altitudine (m)** | 145,46 | 144,67 | 139,38 | 142,72 | 145,50 | 144,51 |
| **Amplitude Altitude (m)** | 87,04 | 12,57 | 22,64 | 33,05 | 51,21 | 51,21 |
| **Min. Slope (grades)** | - | 0,01 | 0,02 | - | - | - |
| **Max. Slope (grades)** | 56,59 | 9,08 | 19,40 | 17,74 | 39,36 | 39,36 |
| **Medium Slope (grades)** | 2,21 | 0,96 | 2,78 | 2,35 | 1,68 | 1,85 |
| **Medium Exposure (grades)** | 203,08 | 190,85 | 195,01 | 206,24 | 217,90 | 213,06 |

## 3. Description of the archaeological site

The actual archaeological site takes up an area of 17.8 km$^2$ (1780,5 ha) and spreads on the three interfluves between the rivers Poştei, Caran, the Lacului river and Valea Viilor (the Vineyard Valley). However, for a better analysis, the entire spatial data base was made for a much wider area, between the settlements of Orţişoara and Seceani to the North and Corneşti and Murani to the South, totalizing 89,35 km$^2$.

The studied area is at the Southern limit of the Vinga Plain (Bizerea 1973), which is a high plane. At its Southern limit it comes into contact with the low, subsiding Timiş Plain. This explains the variation in altitude from 97.6 m in the South and South-East of the studied area to 184.65 m in the North-Eastern part of it.

The relief is made up of smooth, wide interfluves lined up on the East – North-East / West – South – West axis, with medium heights of 140 – 145 m and widths of up to 2600 m, that represent the terraces of the rivers that fragment the plain. The interfluves are separated by the valleys of the rivers Poştei, Caran, Lacului, Valea Viilor and Măgheruş whose valley floors are 20 to 50 m deeper than the interfluves. The valleys have transverse profiles shaped as an open "U", typical of evolved valleys, with the Northern slope more inclined than the Southern one.

The transversal profiles of the valleys are close to the equilibrium profile, except for a sector of the Lacului Valley that present, in the spot in which the second earth wave intersects the valley, a slope rupture of approximately one meter. At the same time there are obvious morphologic differences in the sector upstream from the slope rupture and the one downstream, which suggests the hypothesis that the second earth wave also had the role of a dam at the site of the Lacului Valley. The arguments that support this hypothesis are:

- the existence of the slope rupture of the transversal profile of the valley, situated exactly in the extension of the second earth wave;
- the inexistence of such slope ruptures on the other valleys which suggests the human origin of this one;
- the flat aspect of the riverside upstream from the slope rupture, contrasting with the one downstream from the place in discussion, aspect that suggests the existence of a former sedimentation area (the floor of a lake);
- the high conservation state if the wave on the slopes and towards their base, as compared to the other areas in which the earth waves intersect the water courses.

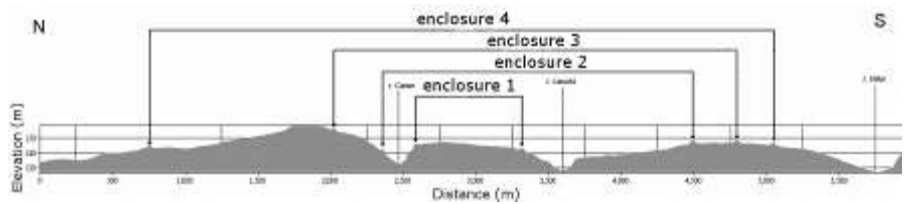

*Fig. 4. Transversal topographic profile on the N-S axis,
between the Poștei and Viilor rivers*

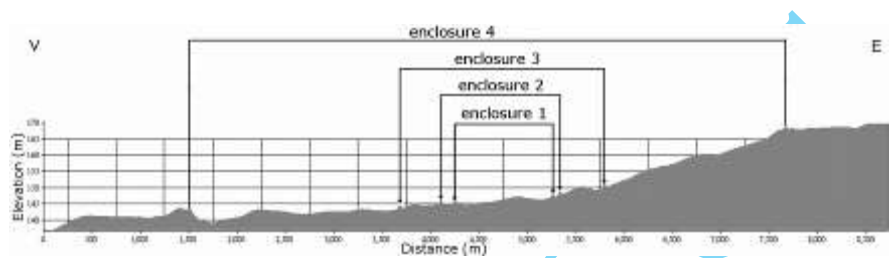

*Fig. 5. Longitudinal topographic profile on the W-E axis,
on the interfluves between the Caran and Lacului rivers.*

## 4. Results

The slope representation (Fig. 6) was generated based on the DEM and enabled the analysis of the declivity values for the entire area and for each enclosure separately (Table 1). Because of the fact that most of the area is taken by the wide and smooth interfluves the medium value of the slope is 2.21, from which the majority consists of the slopes with declivity values below 5 degrees (75 km$^2$). The highest values, of over 10-15 degrees, are characteristic of the valley slopes and take up a surface of only 2 km$^2$. An interesting situation of the distribution of the declivity values is given by the actual presence of the earth waves, which, according to the height and width, lead to the appearance of slopes of more than 5, 10 and even 20 degrees. This is why the earth waves are perfectly visible on the slope representation.

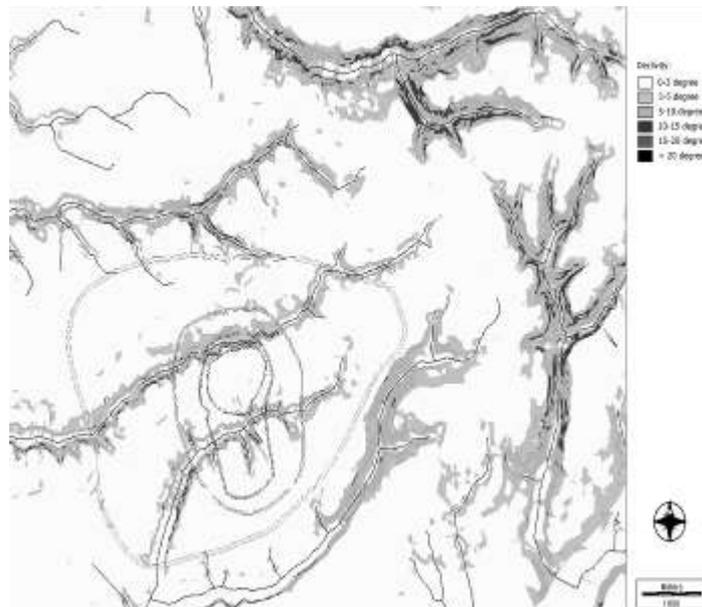

*Fig. 6. The slope representation for the studied area.*

The aspect represents the line on which the slope has its maximum value. It can be defined as "the direction of the horizontal projection of the surface normal of the slope and it is measured clockwise, referring to the direction of the geographical North" (DD98). It is expressed in degrees starting from 0° (the North) to 360° (again North). The flat surfaces, with a 0° slope will be given the -1 value. By reclassifying the aspect values we have the map of the surface exposure (of pixels) referring to the solar radiation on eight directions (Fig. 7).

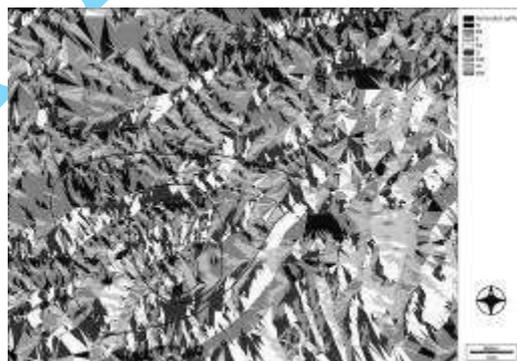

*Fig.7. The exposure map of the surfaces in eight directions*

These can be grouped into four categories: sunlit slopes (S, SW), semi-sunlit slopes (W, SE), semi-shaded slopes (E, NE) and shaded slopes (N, NE). The map is useful for the analysis of the way in which the land was used, the distribution of vegetation and for the analysis of the distribution of the human settlements. Due to the East-West and the North-South lining of the valleys, the share of the surfaces with different orientations is appreciatively equal in the studied area. However, within the archaeological site the situation is different due to the fact that it is crossed only by valleys with a E-W lining and thus, the highest share is that of the surfaces with a N-S exposure. In the case of Enclosure 1 that extends only on the interfluves the areas with a Southern exposure are predominant due to the fact that most of it declines slightly to the South (Fig. 4 and 8).

It is noticeable at the same time that the morphometric parameters of each enclosure, analyzed at a pixel level, although very similar, vary however according to the surface of the enclosures and their positioning on one or more interfluves (Fig. 4, 5 and 8). The biggest differences are between Enclosure 1 and 4.

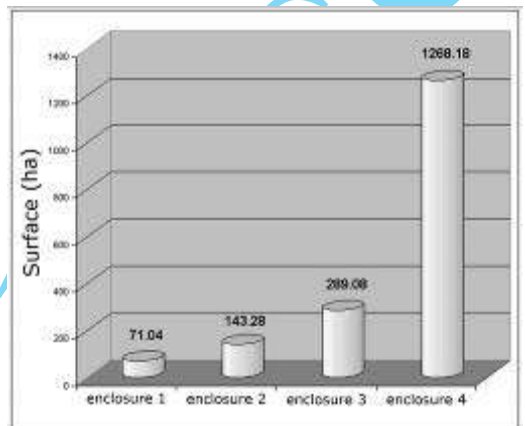

*Fig. 8. The surface of each enclosure*

Enclosure 1, the smallest one (71 ha), lies only on the interfluves between the Caran and Lacului rivers and has the most uniform morphometric values (Table 1). The analysis of the histograms of the morphometric models (Fig. 9) shows that the values are closely grouped around the average. The altitude varies slightly (the amplitude is of only 12 m and the standard deviation is of 1.96) because the surface taken up by Enclosure 1 is an almost flat one (the medium slope is of only 0.96 degrees, the standard deviation = 0.95), slightly inclined towards South (Fig. 4).

Enclosure 4, the largest one (1268,14 ha), lies on all the three interfluves, so it presents the highest variation of the morphometric parameters (Fig. 6) and the highest medium altitude (145.5 m) the altitude varies a lot on the W-E lining, dropping from 168.56 m in the East to 117.35 m in the West, while on the N-S lining the variations are more reduced, from almost 160 m in the North to 130 in the South (Fig. 1 and 2). The amplitude of the altitude values for Enclosure 4 is of 51.21 m and the standard deviation is 9.26 (Fig. 9)

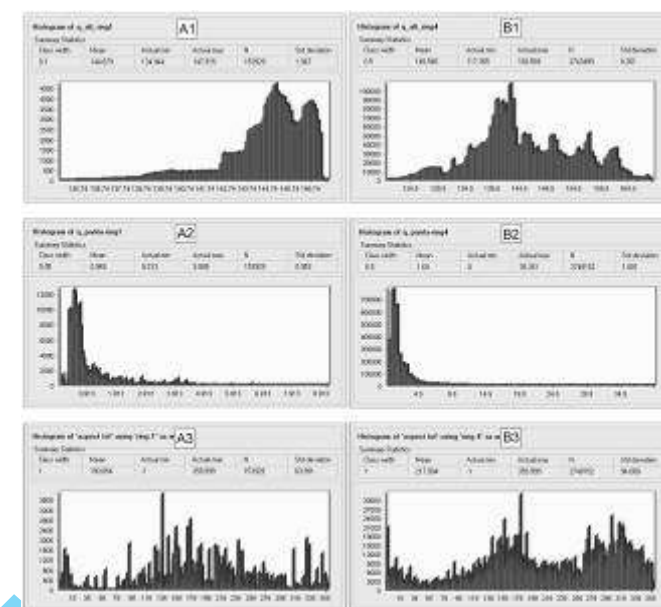

*Fig. 9 The histograms of the morphometric models for enclosures 1 and 4: A1, histogram of the altitudes in Enclosure 1; A2, histogram of the altitudes in Enclosure 1; A3 histogram of the aspect values of Enclosure 1; B1, histogram of the altitudes in Enclosure 4; B2, histogram of the altitudes in Enclosure 4; B3 histogram of the aspect values of Enclosure 4.*

**Conclusions and future work**

Pedologic maps (IPG97) on the scale of 1:10000 were used for the making of digital maps of soils (Fig.10). The maps were scanned and referenced geographically in the national coordinate system Stereo 1970. The analysis of the distribution of the types of soil reveals a clear correlation of these with the relief. As such, in the lowest S-W part there is an area taken up by chernozems, while on the other interfluves there

259

are luvisols that take up most of the surface. The slopes of the valleys are characterized by the presence of regosols while the riversides there ale fluvisols and gleysols.

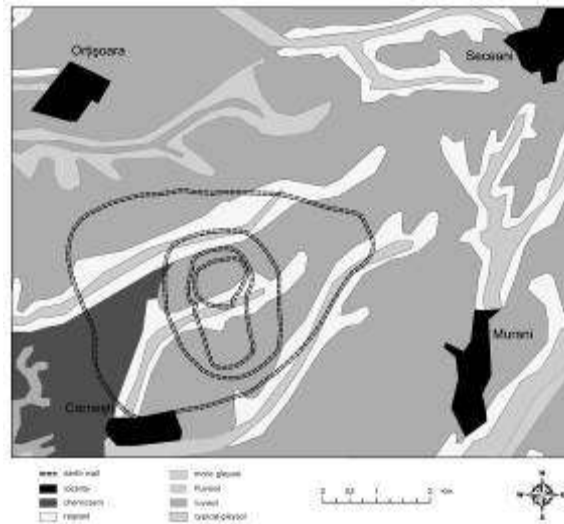

*Fig. 10. Soil map*

As a result of the analysis of these parameters several conclusions with a geo-archaeological character can be drawn. First we can group the results of the research into two main aspects: 1 the elements of habitat, 2. the defense system.

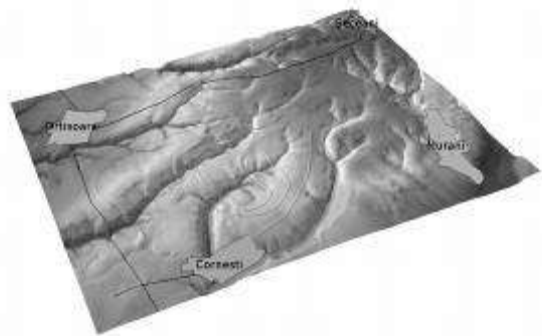

*Fig. 11. Topographic map*

Concerning the habitat, the geomorphologic and cartographic analyses have revealed the positioning of the inhabited spaces in the areas with a S and SW exposure, within enclosures 2 and 3. Another favorable element of habitat, used by the populations located in this area, was the closeness to the water courses whose wide valleys were only slightly inclined and could be dam up and which provided the water volume necessary for a large community.

The transposition of the soil types on an accurate model of the field, provides the possibility to identify the areas affected by human intervention (the habitat and plough land areas − chernozems, the old forest areas, used to maintain the fortification system − luvisols, the area of the possible lake, on the Lacului Valley, created by the dam built by the crossing of the valley by the wave of Enclosure 2 − fluvisols and gleysols).

Concerning the defense system, the analysis of the cartographic model of the land enables the drawing of preliminary conclusions referring to the positioning, the usefulness and the chronological relation among the four fortified enclosures. It can be noticed that, while enclosures 1, 2 and 3 have the role of providing intrinsic protection for the inhabited areas and the areas with a well determined function (see the case of enclosure 1 which can have an elitist/exclusive part or sacred, because up to this day no traces of habitation have been discovered), enclosure 4 has a different part, that of fencing a larger biotic space, which includes the inhabited areas and the subsistence areas (plough land, pastures, forest areas, water and fishing resources).

At the same time, the lining of the 4th fortification wave has a strong military defensive character, following accurately the water-shed lines and including within it the origin of important water sources, vital to a large community that resided within this fortification. The situation from Corneşti "Iarcuri", in which the fortification surrounds a long existing biotic area, is a relatively rare case, which would relate to a sedentary community. It is obvious that this fortification complex is unitary and the four waves of the fortification are contemporary and do not represent the evolution steps of this community in time.

Future exhaustive archaeological digs and interdisciplinary research will complete this preliminary study based on the analysis of satellite images, aerophotograms, cartography and geomorphology of the land on which this archaeological complex is situated.

---

[i] **iarc**, *iárcuri,* s.n. (reg.) large border ditch. In local toponimy "iarc" means a large border ditch, elevation of the ground, earth wave, and it is found as a toponim in other places in Banat region.